\documentclass[aps,prb,twocolumn,superscriptaddress,showpacs,floatfix]{revtex4}
\usepackage{graphicx}
\usepackage{amsmath}
\usepackage{amssymb}
\usepackage{url}

\newcommand{\D}{\ensuremath{{\mathrm{d}}}}

\graphicspath{{.}{../}{./eps/}}

\begin{document}

%\title{Localization properties of quantum random resistor networks}
\title{Diffusion and localization in quantum random resistor networks}
\author{Gerald Schubert}
\author{Holger Fehske}
\affiliation{Institut f\"ur Physik, Ernst-Moritz-Arndt Universit\"at
  Greifswald, 17487 Greifswald, Germany}
%
%\affiliation{Institut f\"ur Physik, Ernst-Moritz-Arndt Universit\"at
%  Greifswald, 17487 Greifswald, Germany}
\date{\today}
\begin{abstract}
The theoretical description of transport in a wide class of novel materials is 
based upon quantum percolation and related random resistor network~(RRN)
models. We examine the localization properties of electronic states of diverse 
two-dimensional quantum percolation models using exact diagonalization in 
combination with kernel polynomial expansion techniques. Employing the local distribution approach we determine the arithmetically and geometrically averaged 
densities of states in order to distinguish extended, current carrying states 
from localized ones. To get further insight into the nature of eigenstates of
RRN models we analyze the probability distribution of the local density of 
states in the whole parameter and energy range. For a recently proposed
RRN representation of graphene sheets we discuss leakage effects.

\end{abstract}

%  72.15.Rn Localization effects (Anderson or weak localization)
%  71.23.An Theories and models; localized states
%  05.60.Gg Quantum transport
%  71.30.+h Metal-insulator transitions and other electronic transitions

\maketitle

\section{Introduction}
Disorder is an intrinsic feature of many solid state materials.
The spatial inhomogeneity of a sample strongly affects the transport
properties, particularly in low-dimensional systems where Anderson
localization effects play an important role~\cite{An58}.
Attempts to describe quantum transport in disordered media
usually rely on free electron models with random links between
lattice sites and/or varying on-site potentials. Then transport is 
related to percolating current patterns in a kind of random resistor 
network (RRN). RRN models apply to both classical and quantum 
percolation problems. For classical percolation the existence
of a sample-spanning cluster ensures finite conductivity above a 
certain percolation threshold of accessible sites $p_c$. 
In the quantum case, the appearance of current carrying states
strongly depends on the (relative) importance of tunneling,
scattering and interference processes. For instance, strong 
scattering at the irregular boundaries of a conducting cluster    
may lead to a localization of the charge carrier even for $p>p_c$.        

Over the past years percolation approaches have been employed in 
many circumstances, e.g., in order to model the transport properties of doped 
semiconductors~\cite{ITA94} and granular metals~\cite{FIS04}, 
the dynamics of atomic Fermi-Bose mixtures~\cite{SKSZL04},
the wave propagation through binary inhomogeneous media~\cite{AL92},
as well as the formation of novel two-dimensional spin liquids~\cite{YRH05}. 
Another focal point is the metal-insulator transition, e.g., in 2D n-GaAs 
heterostructures~\cite{SLHPWR05} and colossal magnetoresistive 
manganite films~\cite{BSLMDSS02}, or rather the superconductor-insulator 
and (integer) quantum Hall transitions~\cite{DMA04,SMK04}. Quite recently,
the problem of disorder in systems with Dirac fermions has been studied in 
the context of dirty superconductors~\cite{HA02}
and two-dimensional (2D) graphene~\cite{PGLPC06,MB07p,XX07,RTB07}.

In mono- and bilayer graphene, and graphene-based field effect transistors, 
strong fluctuations of the local charge density and, thus, the local 
conductivity have been reported near the so-called 
``neutrality point''~\cite{NMMFKZJSG06,MAULSVY08,CF07p,KNG06},
where the conductivity reaches its minimum. 
The mesoscopic regions of different charge carrier density may be caused by
inhomogeneities in the substrate, or non-perfect 
stacking~\cite{TZBZAHDSK07,CJFWI07p,RD08p,KG07p}.
In order to model the minimal conductivity in graphene a RRN representation 
of a graphene sheet has been proposed by Cheianov {\it et al.}~\cite{CFAA07}, 
where random links between electron and hole ``puddles'' 
(corresponding to lattice sites) are assumed to determine the observable 
conductivity rather than the local conductivity of a puddle. 
Such a RRN formulated on a square lattice is closely 
related to a 2D quantum percolation percolation model
with additional finite ``leakage'' between all lattice 
sites. 

Motivated by this situation, in this paper we perform 
an in-depth numerical study of generalized 2D percolative RRN models. 
In particular we analyze how the leakage and connectivity rules   
influence the nature of the single-particle states.
Since in two dimensions the problem of quantum percolation is still discussed 
controversially---especially with respect to the existence of a quantum
percolation threshold $p_c \leq p_q \leq 1$, see, e.g., Refs.~\onlinecite{SG91,MDS95,BKS98,HKS02,OC84,SC84,LZ99,DCA00,ERS01,NBR02,IN07p,SF08}---we rely on unbiased numerical 
techniques, which take quantum effects fully into account. 
To this end we employ the so-called local distribution~(LD) 
approach~\cite{AF06,AF08}. This technique, based on the determination 
of the distribution of the local density of states (LDOS) for all energies, 
has been previously applied to tackle localization phenomena 
in various disordered systems with great success~\cite{DPN03,SWWF05,ASWBF05,SWF05b,DK97,BHV05,BF02,BAF04}. 

The paper is organized as follows. Sect.~II introduces the RRN model
and briefly outlines the LD approach. In Sect.~III, we examine the 
localization properties of the eigenstates of RRN models 
by calculating the LDOS distribution, as well as the 
average and typical density of states. Sect.~IV contains our conclusions.   
\section{Model and Methods}

\begin{figure}[t]
  \centering  
  \includegraphics[width=\linewidth,clip]{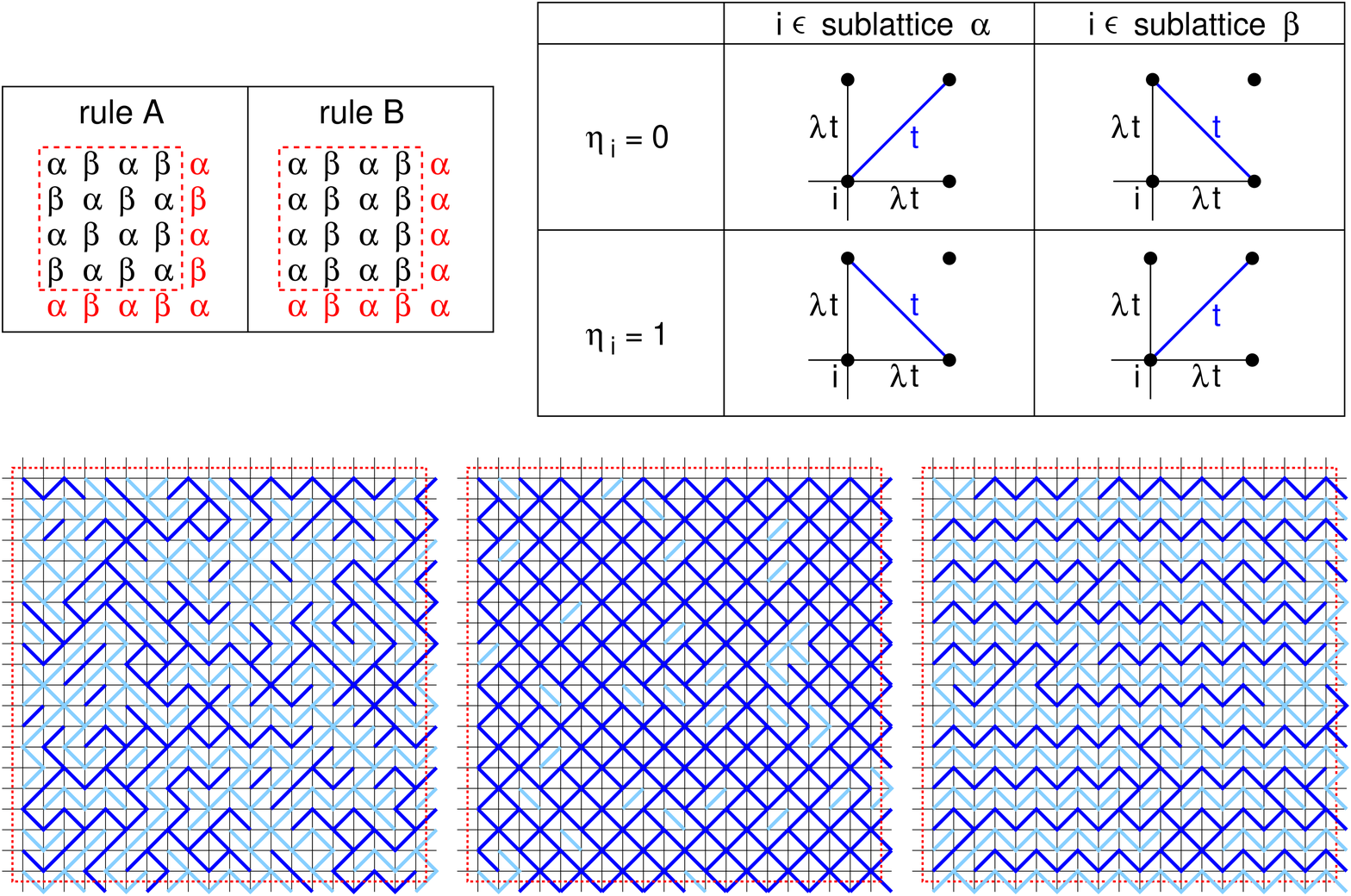}
 
  \caption{(Color online) Upper left panel: Generation rules A and B for
    sub-lattices $\alpha$ and $\beta$. Upper right panel:
    Connection rule of neighboring sites on different sub-lattices.
    Lower panels: Three particular cluster realizations representing the above
    rules on a $20\times20$ lattice. From left to right: $p=0.50$,
    $p=0.90$ (rule A), $p=0.90$ (rule B). Note that 
    rules A and B become equivalent for $p=0.50$.
    At the red dashed lines the system is interconnected by 
    periodic boundary conditions.}
  \label{fig:RRN_model}
\end{figure}

Let us consider a 2D square lattice with $N$ sites on two sub-lattices 
$\alpha$ and $\beta$, which, e.g., might represent regions of different charge 
carrier concentrations. The sub-lattices are linked by connection 
rules characteristic of specific materials. Generation rule A 
corresponds to a checkerboard-like structure (see 
upper left panel of Fig.~\ref{fig:RRN_model}). 
Regions (sites) of the sub-lattices are randomly connected with 
each other (lower panels of Fig.~\ref{fig:RRN_model}) according to the 
connection rules displayed in the upper right panel 
of Fig.~\ref{fig:RRN_model}. The hopping probability between 
suchlike connected sites is assumed to be much higher 
than for hopping events to nearest neighbors.
The latter ones are reduced by the leakage $\lambda<1$.

For the more abstract case of generation rule B 
(with the above connection rule), additional random bonds 
connect $\alpha$ to $\beta$ sites, favoring the formation of 
quasi one-dimensional zig-zag chains. We still have a weak 
leakage between all neighboring sites. This generation rule 
can be interpreted as an attempt to incorporate a spatial
anisotropy into the model.

In both cases, the corresponding RRN can be described by the Hamiltonian
\begin{equation}
  \begin{split}
    H   =  -t\Big[
      &  \sum \limits_{i\in \alpha} \left(\eta_i c^{\dag}_i c_{i+\nearrow}   
        + (1-\eta_i) c^{\dag}_{i+\uparrow} c_{i+\rightarrow}\right)  \\
      &  + \sum \limits_{i\in \beta}  
      \left( \eta_i c^{\dag}_{i+\uparrow} c_{i+\rightarrow}   
        + (1-\eta_i)c^{\dag}_i c_{i+\nearrow} \right)  \\
      & + \lambda  \sum \limits_{i} \left( c^{\dag}_i c_{i+\rightarrow}   
        c^{\dag}_i c_{i+\uparrow} \right)\Big] + \text{H.c.}\;.
  \end{split}
\label{model_RRN}
\end{equation}
Here $c^{\dag}_i$ ($c^{}_i$) creates (annihilates) an electron at 
site $i$ and the arrows denote the nearest neighbors site of $i$ 
in the corresponding direction. The discrete random variables 
$\eta_i \in \{0,1\}$ determine which diagonal in the plaquette is 
connected (cf. upper right panel in Fig.~\ref{fig:RRN_model}). 
By fixing the expectation value of the $\{\eta_i\}$--distribution, 
$p=\langle\eta_i\rangle$, we can control the sizes of the regions 
of connected lattice sites. For $p=0.5$ the RRN model~(\ref{model_RRN}) 
does not depend on the generation rules A or B because of the symmetry 
of the bimodal $\{\eta_i\}$--distribution. 
In the limit of vanishing leakage we obtain the standard 2D quantum 
percolation model using generation rule A.

To characterize the transport behavior of our model we adopt
the local distribution approach (for a detailed description of 
this technique, we refer the reader to Ref.~\onlinecite{SF07p} 
and references therein). The main idea is the following: 
We calculate the LDOS for all lattice sites $i$,
\begin{equation} \label{LDOS}
  \rho_i(E) = \sum\limits_{n=1}^{N}
  | \langle i | n \rangle |^2\, \delta(E-E_n)\;.
\end{equation} 
This quantity depends on the energy and varies from site to site.
Moreover the $\rho_i$ are specific for a particular sample 
(realization $\{\eta_i\}$). Calculating and recording $\rho_i$ 
for many sites and realizations we obtain the probability 
distribution $f[\rho_i]$ and distribution function
\begin{equation}
  F[\rho_i]=\int_0^{\rho_i} f[\rho_i']\,\D \rho_i'\;.
\end{equation}
of $\{\rho_i\}$. Both $f[\rho_i]$ and $F[\rho_i]$ are
self-averaging. That is, in the thermodynamic limit, they are   
independent of the actual realization $\{\eta_i\}$ and the chosen 
sites $\{i\}$, but will be solely determined by the (global) model 
parameters $p$, $\lambda$, and the generation rule. 
This restores translational invariance on the level of distributions.

A well-established criterion for localization then can be deduced from
the shape of the distribution function. 
Since for an extended (current carrying) state the amplitude of 
the wave function is more or less uniform, $F[\rho_i]$ steeply 
rises in the vicinity of the mean value of the LDOS, 
$\rho_{\text{me}}=\langle \rho_i \rangle$. For localized states,
on the other hand, the LDOS strongly fluctuates throughout the lattice 
leading to a rather gradual increase of $F[\rho_i]$ as a function of the magnitude of
$\rho_i$.  To capture this different behavior quantitatively, we may compare the 
arithmetic mean of the LDOS, $\rho_{\text{me}}$,
to its the geometric mean, the so-called ``typical'' DOS, 
$\rho_{\text{ty}}= e^{ \langle \ln (\rho_i)\rangle}$, see, e.g.,
Ref.~\onlinecite{SWWF05}.

For extended states $\rho_{\text{me}}$ and $\rho_{\text{ty}}$ are of the same
order, whereas $\rho_{\text{ty}}$ vanishes for a localized state
(or, in a finite system, is at least considerably suppressed).  
Of course, a reliable distinction between extended and localized states
requires the consideration of different system sizes. For extended 
states $F[\rho_i]$ is independent of $N$. By contrast, for localized 
states $F[\rho_i]$ shifts toward lower values as $N$ increases  
(causing the reduced value of $\rho_{\text{ty}}$).

The LDOS can be calculated very efficiently by means of the kernel
polynomial method (KPM)~\cite{WWAF06}. Within this technique, 
the spectrum of $H$ is expanded into a finite series
of Chebyshev polynomials with additional damping factors.
This approximation can be viewed as a convolution of the spectrum
with the Jackson kernel, an almost Gaussian of width $\sigma$.
As $\sigma$ depends on the order of the Chebyshev series as well as on the
energy, we have to adapt the expansion order to ensure an uniform resolution  
(constant $\sigma$) for the whole spectrum~\cite{SF07p}.
\section{Results}
\begin{figure}[b]
  \centering 
  \includegraphics[width=\linewidth,clip]{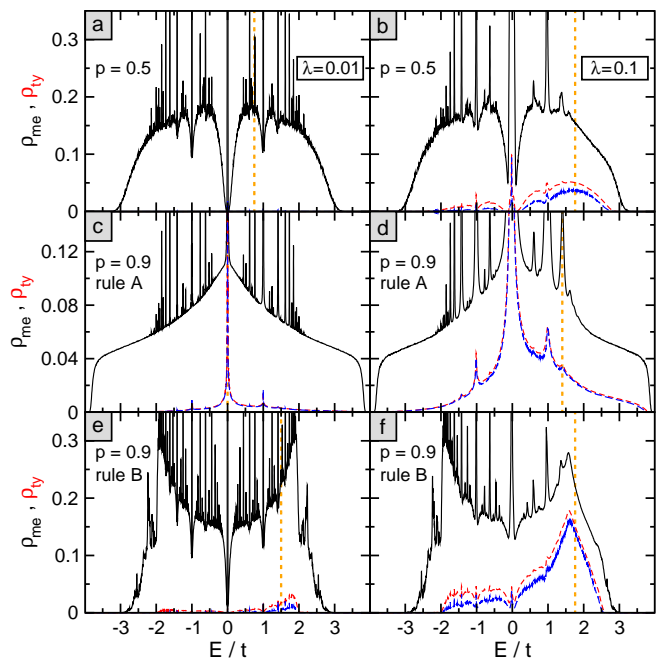}
   \caption{(Color online) Mean (upper solid lines) and 
     typical (lower dashed lines)
     DOS. Results are shown for $N=400^2$ and $N\sigma=45$ for different 
     $p$, $\lambda$ and generation rules A and B.
     To illustrate the finite-size dependence of $\rho_{\text{ty}}$
     the results for a $N=800^2$ lattice are included 
     (long dashed lines). The data are based on $10^5$ $(10^4)$ 
     individual LDOS spectra for $N=400^2$ $(800^2)$.
     Dotted vertical lines indicate the energies for which the
     distribution functions of the LDOS are given in Fig.~\ref{fig:DistFkt}.}
   \label{fig:RRN_mety}
\end{figure}
%
% Consider rho_me
%
In Fig.~\ref{fig:RRN_mety} we compare the mean and typical DOS
in dependence on $\lambda$ and $p$ for generation rules A and B. 
While the 2D percolation model ($\lambda=0$, rule A)
exhibits symmetric DOS spectra, the inclusion of next-nearest 
neighbor hopping causes an asymmetry that grows with increasing $\lambda$.
Moreover, if $p>0.5$, the results obtained for the two generation rules 
differ significantly. For the A-case the mean DOS resembles the 2D DOS.
Generation rule B leads to a spectrum evocative of the DOS 
in 1D. In both cases a multitude of spikes exists which 
we can attribute to localized states on ``isolated'' islands. 
This feature is well known from the binary alloy model~\cite{SWF05}.
Increasing the leakage broadens the peaks and reduces their abundance, 
because the ``isolation'' of the islands is weakened.
Furthermore, the leakage shifts the ``special energies'' at which these
peaks appear toward lower values as compared to those in the 
standard 2D percolation model~\cite{SWF05}.

The typical DOS underlines the prominent role of the
leakage. In Fig.~\ref{fig:RRN_mety} (a) a vanishing $\rho_{\text{ty}}$ 
suggests that all states are localized. 
As long as the leakage is very small, the typical DOS remains 
small as well throughout the band. This also holds
for larger $p$ and both generation rules. 
Sizeable values of $\rho_{\text{ty}}$ appear at larger values of
$\lambda$ only (compare the data for $\lambda=0.01$ and 0.1). 
As for the mean DOS, 
we observe a pronounced asymmetry of $\rho_{\text{ty}}$
in the electronic band. This favors a finite typical DOS, 
i.e. extended states, for $E>0$. However, this has to be taken with
caution. If we increase the system size together with the 
resolution of the KPM in such a way, that the number of states 
within the width of the Jackson kernel, $N\sigma$, is 
kept constant, the picture changes. While for truly extended states, 
$\rho_{\text{ty}}$ should be independent of the
system size (cf. Fig.~\ref{fig:RRN_mety} (d)),
in Figs.~\ref{fig:RRN_mety} (b) and (f), the typical 
DOS decreases with increasing system size. 
This points towards a large localization length. 
While for small and moderate $N$ the state still spans the 
entire lattice,  the system size exceeds the localization length
for larger $N$. Therefore, on a considerable number of sites the 
LDOS is very small, leading to a reduced value of $\rho_{\text{ty}}$. 
Then the question arises, why in panel Fig.~\ref{fig:RRN_mety} (d)
the typical DOS is reduced as compared to $\rho_{\text{me}}$, even though 
this ratio is independent of $N$. This can be easily understood if we  
consider a perfectly ordered system, i.e. $p=1$.
There the sites which do not belong to the spanning sub-lattice are 
also taken into account, and amount to half of the lattice sites.
In absence of leakage, at those sites the LDOS vanishes and probing 
such a site completely suppresses $\rho_{\text{ty}}$. 
For finite $\lambda$, however, these sites have small amplitudes
and in this manner $\rho_{\text{ty}}$ is reduced as compared 
to $\rho_{\text{me}}$ but finite. These arguments also hold for 
panel Fig.~\ref{fig:RRN_mety} (b), but there
the low leakage suppresses the typical DOS already almost completely, 
except for $E=0$.

As mentioned above,  the comparison of $\rho_{\text{ty}}$ and $\rho_{\text{me}}$
may only serve as a first indication of localization. 
A careful finite-size scaling of the full probability function
(or the distribution function) of the LDOS allows for a more
reliable distinction between extended and localized states.
%
%%%%%%%%%%%%%%%
%
% Zu Fig. 3
%
%%%%%%%%%%%%%%
\begin{figure}[t]
  \centering  
  \includegraphics[width=\linewidth,clip]{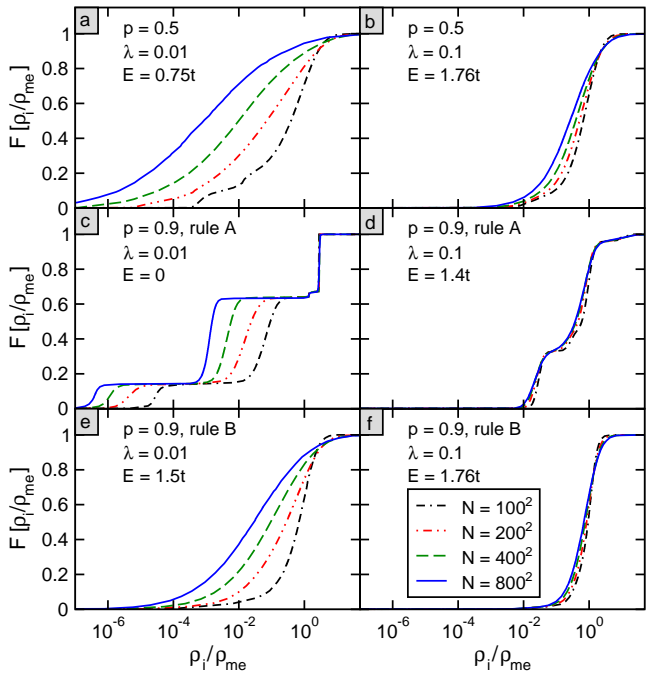}
  \caption{(Color online) Distribution function of the normalized LDOS for 
    the energies indicated by vertical lines in Fig.~\ref{fig:RRN_mety}.
    The statistics is based on $10^7$, $10^6$, $10^5$, $10^4$ LDOS values for
    $N=100^2, 200^2, 400^2, 800^2$ and resolution $N\sigma=45$.}
  \label{fig:DistFkt}
\end{figure} 
Thereto, in Fig.~\ref{fig:DistFkt} we show for the same sets of parameters
the distribution function of the LDOS at various characteristic energies. 
In the panels of Fig.~\ref{fig:RRN_mety}  these energies are indicated
by vertical dotted lines. 

Here one crucial point concerns the normalization 
of $\rho_i$ to $\rho_{\text{me}}$. 
For a completely extended state, with uniform amplitudes,
the distribution function would be a step function. 
Due to the used scaling, this step would be located at $\rho_i/\rho_{\text{me}}=1$, 
irrespective of the system size. The random structure of the underlying 
lattice will distort this perfect jump-like behavior of $F[\rho_i/\rho_{\text{me}}]$.
But nevertheless extended states are characterized by an  
$N$-independent distribution function [cf. Fig.~\ref{fig:DistFkt} (d)]. 
In contrast, for localized states, 
$F[\rho_i/\rho_{\text{me}}]$ constantly shifts 
toward lower values as $N$ increases [cf. Figs.~\ref{fig:DistFkt} (a), (e)]. 
Depending on the localization length, this effect is more or less pronounced. 
Thus, a wide range of system sizes is necessary in order to discriminate 
localized from extended states by means of finite-size scaling [cf. 
Figs.~\ref{fig:DistFkt} (b) and (f)].

A particular interesting shape of $F[\rho_i/\rho_{\text{me}}]$ appears 
in Fig.~\ref{fig:DistFkt} (c), for a state in the band center. 
From the multi-step structure, we may deduce that the probability 
distribution is mainly concentrated around three values. 
The largest of those values is independent on $N$, while the others
show the above discussed variation. This behavior can be explained 
by considering again the completely ordered case. In absence of leakage 
the $E=0$ eigenstate is highly degenerate: we have 
$N/2$ completely localized states, one on each isolated site. 
The other half of the eigenstates are extended in the perfect 
lattice, providing energies in the whole band. 
As the LDOS probes the complete eigenspace and not just the amplitude 
of one particular eigenstate, at $E=0$ we obtain the same value of 
$\rho_i/\rho_{\text{me}}$ for each isolated site. 
Introducing a small leakage kills the high degeneracy in principle.
Due to the finite energy resolution of the KPM, however, still many 
of these states contribute to the LDOS at $E=0$. 
In the presence of imperfections ($p<1$), some of the former 
isolated sites will now be connected to form larger islands.
Then less than $N/2$ states will contribute to the ``isolated islands'' peak 
at high $\rho_i/\rho_{\text{me}}$  [but still around 38\% 
in Fig.~\ref{fig:DistFkt} (c)]. 
The second step, having a weight of about 50\%, is due to sites on the 
highly connected majority sub-lattice, on which the LDOS is reduced 
but finite due to the leakage. The remaining lowest step originates 
from more complicated islands of several sites. On those islands 
($2,3,\ldots$ sites), the $E=0$ eigenstates have vanishing
amplitudes on some sites. Due to the leakage these sites again 
acquire a finite value of $\rho_i$. 
%While we are able to explain the structure of $F[\rho_i/\rho_{\text{me}}]$
%for $E=0$  [Fig.~\ref{fig:DistFkt} (c)], for general parameters,
%the situation is more difficult.

%%%%%%%%%%%%%%%
%
% Zu Fig. 4
%
%%%%%%%%%%%%%%

\begin{figure}
  \centering 
  \includegraphics[width=\linewidth,clip]{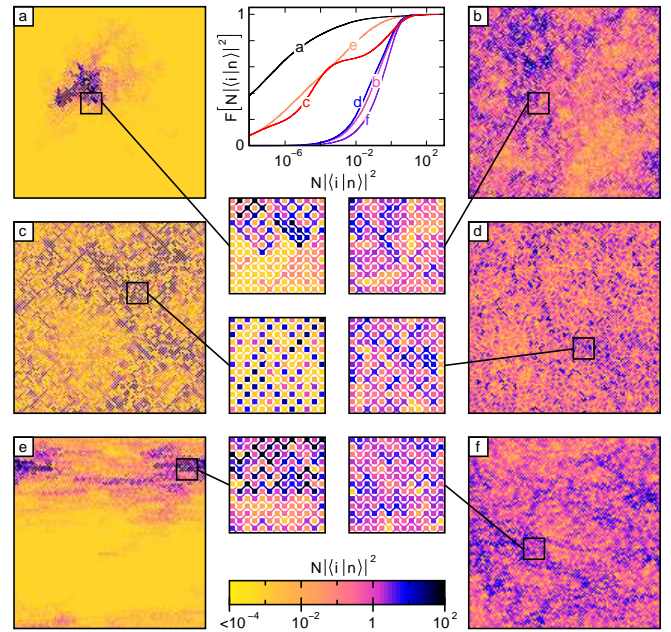}
   \caption{(Color online) Normalized occupation probability
     $N|\langle i| n\rangle|^2$  of several characteristic 
     eigenstates $|n\rangle$ on a $N=128^2$ lattice. 
     The generation rules, doping and leakage 
     as well as the chosen energies $E_n$ correspond to the ones 
     for which the distribution functions of the LDOS has been presented in 
     Fig.~\ref{fig:DistFkt} (same order of panels). In the center column 
     the distribution function of the normalized occupation probability
     is shown for these states. Furthermore, for each state an enlargement
     of a characteristic region together with the local structure of the 
     present links is shown.
   }\label{fig:CharState}
\end{figure}

To get additional insight into the spatial structure of the eigenstates, we 
investigate a smaller $128\times128$ system using exact diagonalization techniques. 
Figure~\ref{fig:CharState} visualizes characteristic eigenstates  
for the parameters discussed in Figs.~\ref{fig:RRN_mety} and~\ref{fig:DistFkt}. 
Most notably is the 
pronounced leakage-dependence of the results. 
While we see clear localization for $\lambda=0.01$ 
[Fig.~\ref{fig:CharState} (a), (e)], in the high-leakage 
case all states span the entire lattice. The 1D structure for rule B 
and large $p$, which we already found in the DOS in 
Fig.~\ref{fig:RRN_mety} (e), is also predominant in Fig.~\ref{fig:CharState} (e).
In Fig.~\ref{fig:CharState} (c), notably in the magnifying inset, we see 
our assumption confirmed, that %the majority sub-lattice has only very small 
%amplitudes. Sites 
sites
with large amplitudes are almost exclusively 
isolated, with the exception of some larger islands. 
Having the sharp step in mind, which occurs in Fig.~\ref{fig:DistFkt} (c) 
at large $\rho_i/\rho_{\text{me}}$, the alert reader may wonder about 
the different amplitudes in Fig.~\ref{fig:CharState} (c) and the 
rather smooth increase of $F[N|\langle i| n\rangle|^2]$ in the middle column. 
This discrepancy is again due to the high degeneracy of the $E=0$ state and the
fact, that the LDOS takes into account the whole $E=0$ eigenspace.
%
%But for $N|\langle i| n\rangle|^2$, we just randomly pick one particular
In contrast, for $N|\langle i| n\rangle|^2$, we randomly pick one particular
eigenstate out of this subspace.  For the other $E=0$ eigenstates, 
$F[N|\langle i| n\rangle|^2]$ looks similar, only the location of the sites with 
maximum amplitudes differ. Clearly, the physically relevant quantity is the LDOS, 
as the choice of eigenvectors which span the eigenspace is arbitrary up to 
any linear combinations of them. 
Summing up all eigenstates to $E=0$ for a fixed lattice site 
we indeed get the same amplitude on each isolated site, which brings 
us back to the findings of Fig.~\ref{fig:DistFkt}~(c).
Finally we consider the case of higher leakage ($\lambda=0.1$), where all 
states look rather similar. Here the amplitudes fluctuate over a 
smaller range, and the fluctuations occur on very short spatial scales. 
This is most pronounced in Figs.~\ref{fig:CharState} (d), (f) where no 
global structures can be distinguished, in accordance with the 
notion of extended states. 
By contrast, although the state in Fig.~\ref{fig:CharState} (b) spans the
entire lattice, we observe sizeable regions with higher and lower 
amplitudes than the mean value. Presumably, these inhomogeneities 
are even more pronounced for larger systems which, however, 
(as yet) are not accessible by our exact diagonalization studies.
In any case, the shifting of the LDOS distribution function in 
Fig.~\ref{fig:DistFkt} (b) suggests this state to be localized on 
large length scales.

\section{Summary}
In this work we investigated the two-dimensional quantum percolation 
problem for a broader class of random resistor network models 
including leakage terms. Combining exact diagonalization, Chebyshev expansion
and local distribution approaches in calculating the local 
density of states, we determine---after a careful
finite-size scaling---the localization properties of the 
single-particle eigenstates. We found that current carrying states exist, 
and that the appearrance of diffusion 
is mainly triggered by the amount of leakage. 
In contrast to previous work we analyzed the nature of the single-particle
states in the whole energy band and not just the vicinity of 
the neutrality point in the band center $E=0$. 
A tendency toward extended states is observed for $E>0$. 
In view of the simplicity of the considered RRN model, 
our data are certainly not yet suited to be compared 
against real transport data for, e.g., undoped graphene monolayers. 
Nevertheless, using unbiased numerics on a microscopic level, 
we fully account for quantum effects in a RRN model originally 
designed to describe transport in a graphene-based field effect transistor, 
and received results that support a minimal conductivity in graphene.

\section*{Acknowledgments}
The numerical calculations have been performed on the HLRB at LRZ
Munich and the TeraFlop compute cluster at the Institute of
Physics, Greifswald University.

%\bibliography{./ref} 

%\bibliographystyle{apsrev}

\end{document}